\documentstyle[prb,aps,preprint]{revtex}

\begin{document}
\title{Magnetic behavior of single crystalline Ho$_2$PdSi$_3$}
\author{E.V. Sampathkumaran,$^{1,*}$ H. Bitterlich$^2$  K. K. Iyer$^1$,    W. L\"oser$^2$ and  G. Behr$^2$ }

\address{$^1$Tata Institute of Fundamental Research, Homi Bhabha Road, Colaba,
Mumbai - 400005, India}
\address{$^2$Institut f\"ur Festk\"orper- und Werkstofforschung Dresden, Postfach 270016, D-01171 Dresden, Germany}
\maketitle

\begin{abstract}The magnetic behavior of single-crystal Ho$_2$PdSi$_3$, crystallizing in an AlB$_2$-derived hexagonal structure, is investigated by magnetic susceptibility ($\chi$) and electrical resistivity ($\rho$) measurements along two directions.    There is no dramatic anisotropy in the high temperature Curie-Weiss parameter or   in the $\rho$ and isothermal magnetization data, though there is a noticeable anisotropy in the magnitude of $\rho$ between two perpendicular orientations. The degree of anisotropy is overall less prominent than in the  Gd (which is an S-state ion!) and Tb analogues. A point of emphasis is that this compound undergoes long range magnetic ordering below 8 K as in the case of analogous Gd and Dy compounds. Considering this fact for  these compounds with well-localised f-orbital,  the spin glass freezing noted for isomorphous U compounds in the recent literature could be attributed to the role of the f-ligand hybridization, rather than just Pd-Si disorder.
\end{abstract}
\vskip0.5cm
$^*$Corresponding author. E-mail address: sampath@tifr.res.in
\vskip1cm
\newpage
\maketitle
The investigation of intermetallic compounds of the general formula, R$_2$TX$_3$ (R= rare-earth or U, T = transition metal and X = Si  or Ge), derived from AlB$_2$-type hexagonal structure [1], is of great interest, considering that the properties of many of these compounds are relevant  to current problems in the field of magnetism and superconductivity [see, for instance,  references 1-12]. We have also carried out intense studies on the single crystalline forms of the series R$_2$PdSi$_3$ for R= Ce, Gd, and Tb, and found novel anisotropic behavior in magnetic and transport data [5-7]. It was however  noticed that such  anisotropy is less evident in the case of Dy   [8],   and this finding is surprising considering that even the Gd (which is a S-state ion) compound exhibits strongly anisotropic properties. In view of this situation, we considered it worthwhile to extend single crystal investigations to other members of this series. Therefore, we have  investigated  single crystalline Ho$_2$PdSi$_3$, the results of which are reported in this article. Another motivation to take up this investigation is that the analogous U compounds have been found [12] to undergo spin glass freezing,  whereas previous reports [1, 5,  8,  10, 11]  for the above heavy rare-earth compounds reveal features due to long range magnetic ordering.  It is therefore of interest to carefully investigate single crystals of additional f-localised systems, in order to explore possible role of f-hybridization to decide the spin glass ground state in U compounds.  

The single crystals of Ho$_2$PdSi$_3$  (25 to 50 mm in length and 7mm in diameter) were grown by the floating zone method by rf inductive heating from stoichiometric polycrystalline feed rods. The orientation of the single crystals was determined  by  x-ray Laue backscattering. Pieces ( 2mm X 2mm X 10 mm) with two orientations [10$\bar{1}$0] and [0001] were cut from this rod for measurements. Isothermal magnetization (M) measurements at 2 and 5 K and magnetic susceptibility ($\chi$) measurements in the temperature (T) interval 1.8 - 300 K were performed employing a commercial superconducting quantum interference device (Quantum Design) as well as a vibrating sample magnetometer (Oxford Instruments). The electrical resistivity ($\rho$) measurements were performed by a conventional four-probe method employing silver paint for  electrical contacts. 

The T dependence of $\chi$ at low temperatures taken at 100 Oe and 2 kOe for the zero-field-cooled (ZFC) condition of the crystal is shown in Fig. 1. For H= 100 Oe, we have in addition taken the data for the field-cooled (FC) state of the specimen. It is distinctly clear from this figure that, for H//[0001], there is a well-defined peak at 8 K, establishing the onset of magnetic ordering. The transition temperature is  higher than that reported (6 K) for polycrystals [13]. It should be noted that there is no separation of ZFC-FC curves as in the case of  single crystals of Gd, Tb and Dy alloys [5,7,8], which is a sufficient proof that this  compound undergoes  long-range magnetic ordering, rather than spin-glass freezing. As the T is lowered, there is an upturn in $\chi$ values below 4 K, which is indicative of possible spin reorientation effects apparently  characteristic of this class of compounds [5,7,8]. If  $\chi$ data are collected at H= 2 kOe, this upturn vanishes; however, a broad shoulder appears around 4 K in the plot of $\chi$ versus T as though there is a change in the magnetic structure with decreasing T. Now turning to the data for H//[10$\bar{1}$0], the 8 K transition appears only as a broad shoulder in the plot of $\chi$ versus T; however, the transition below 4 K appears as a prominent feature, that is, a tendency to flatten for both the fields. The $\chi$ values are noticeably lower for this direction  compared to those for H//[0001].   In short, $\chi$ behavior appears to be anisotropic in the magnetically ordered region. 

With respect to the paramagnetic  behavior (Fig. 2), the plot of inverse $\chi$  versus T is nearly linear, though there is a tendency for a small deviation below about 50 K presumably due to crystal field effects. The nature of the deviation is different for these orientations (see Fig. 2) in the sense that the experimental values at low temperatures in the paramagnetic state (say, in the range 10 - 30 K) are higher for  H//[10$\bar{1}$0]  with respect to the straight line extrapolated from the high temperature ($>$100 K) linear region, whereas for H//[0001] the values are lower.  The value of  the effective moment obtained from the high temperature linear region is very close to that expected for trivalent Ho ions (close to 10.6 $\mu_B$/Ho mol) and the Curie-Weiss temperature ($\theta$$_p$) is practically the same   for both  directions (close to 2 K). Thus, the anisotropy is insignificant in the high temperature $\chi$ data. It is to be remarked that, the absolute value of  $\theta$$_p$ is nearly the same as the N\'eel temperature (T$_N$) for H//[0001], as inferred from the linear region below 30 K; however the sign of 
$\theta$$_p$  is  positive,  implying the existence of significant ferromagnetic correlations. The low temperature $\theta$$_p$  however changes sign (to -3 K) for H//[10$\bar{1}$0] consistent with antiferromagnetic coupling. 

The isothermal M behavior (Fig. 3) shows no hysteresis  at 2 or 5 K. A close inspection of the data at 5 K suggests that there is a sudden increase in the slope in a field of about 2 kOe, as though there is a spin reorientation, for both the directions; the field at which this occurs increases to about 3 kOe at 2 K as shown in an expanded form in the inset for one direction. This finding suggests that the zero-field magnetic ordering is antiferromagnetic. The magnetic moment at high fields tends towards saturation to a constant value, which is indicative of field-induced ferromagnetism. The value of the saturation moment is somewhat lower than that expected for trivalent Ho ions (10 $\mu_B$/Ho ion) and it appears that higher fields are required to attain this saturation value. It is to be noted that, for the 2 K data for H//[0001], the values at high fields are comparatively lower  than those at 5 K, which may be a consequence of the change in spin orientation as the T is lowered.  Needless to remark that, isothermal M in the paramagnetic state is a linear function of H (not shown in the figure).    

In Fig. 4, we show the T dependence of $\rho$. The low T data has also been shown in an expanded form in the insets. Above T$_N$,  $\rho$ for both crystallographic directions of excitation current has a positive temperature coefficient typical of metals (as expected). Though the values for  the two directions  are  different, the features are qualitatively the same, even below T$_N$. The value of  $\rho$, instead of exhibiting a drop at T$_N$ due to the loss of spin-disorder contribution, shows an upturn. This observation signals the formation of magnetic Brillouin-zone boundary gaps below 8 K, thereby endorsing the previous conclusion from neutron diffraction data [3] that the magnetic structure is of a modulated type (incommensurate with the lattice).

To conclude, we have reported magnetic behavior of single crystalline Ho$_2$PdSi$_3$ for two orientations. Among various findings, we would like to emphasize that the overall features  in isothermal M, temperature dependence of $\rho$ and paramagnetic $\chi$ behavior are essentially isotropic, though the shapes of the $\chi$  versus T plots in the magnetically ordered state appear to show some anisotropy and the magnitude of $\rho$ is noticeably different for  two perpendicular orientations. In this sense, the properties of this compound are similar to that of Dy$_2$PdSi$_3$ [8]. In sharp contrast to this situation,  the features in isothermal M and $\rho$  near the magnetic region are strongly anisotropic not only in the Tb analog but also for the Gd (which is an S-state ion) analog [5,7].  Finally, with  the observation of long range magnetic ordering in another heavy rare-earth compound of this structure in single crystalline form, it can be stated that  the spin-glass behavior of the isostructural U compounds is determined in some fashion by f-hybridization (endorsing the conclusion in Ref. 12), rather than Pd-Si disorder alone, with the assumption that  the degree of possible Pd-Si site disorder is the same for U as well as for the heavy rare-earth members of this series with long range magnetic order.  

Some of the authors (H.B., W.L. and G.B) gratefully acknowledge the financial support of DFG by the Sonderforschungsbereich 463 'Rare Earth-Transition-Metal-intermetallic compounds: Structure, Magnetism, Transport'.

\newpage
\begin{figure}
\caption{The low temperature magnetic susceptibility behavior of Ho$_2$PdSi$_3$ for two crystallographic orientations measured in magnetic fields of 100 Oe and 2 kOe. ZFC and FC represent zero-field-cooled and field-cooled state of the specimens to 1.8 K and the curves corresponding to these two situations overlap.  The vertical arrow  shows the onset of magnetic ordering appearing as a shoulder in the H= 100 Oe data.}
\end{figure}
\begin{figure}
\caption{Inverse susceptibility as a function of temperature for two crystallographic  orientations of  Ho$_2$PdSi$_3$ in a  magnetic field of 2 kOe. The lines represent least squares fitting of the data above 100 K.}
\end{figure}
\begin{figure}
\caption{Isothermal magnetization data at 2 and 5 K  for two crystallographic orientations of  Ho$_2$PdSi$_3$. Low-field behavior is shown in the inset to highlight the existence of a metamagnetic-like transition.}
\end{figure}
\begin{figure}
\caption{Electrical resistivity as a function of temperature for  Ho$_2$PdSi$_3$ for two different crystallographic directions. The low temperature behavior is shown in expanded form as insets to highlight the existence of an upturn.}
\end{figure}

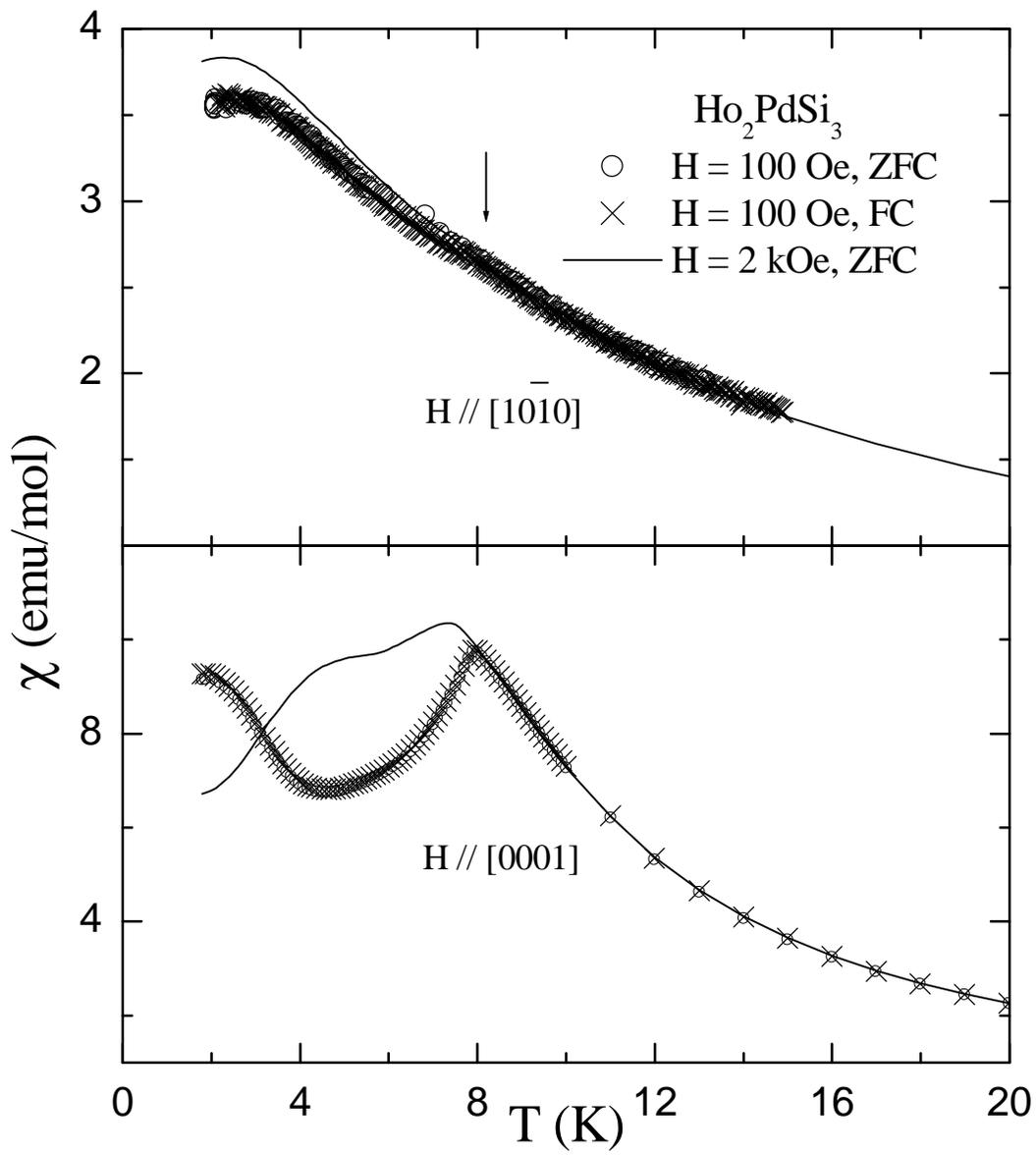
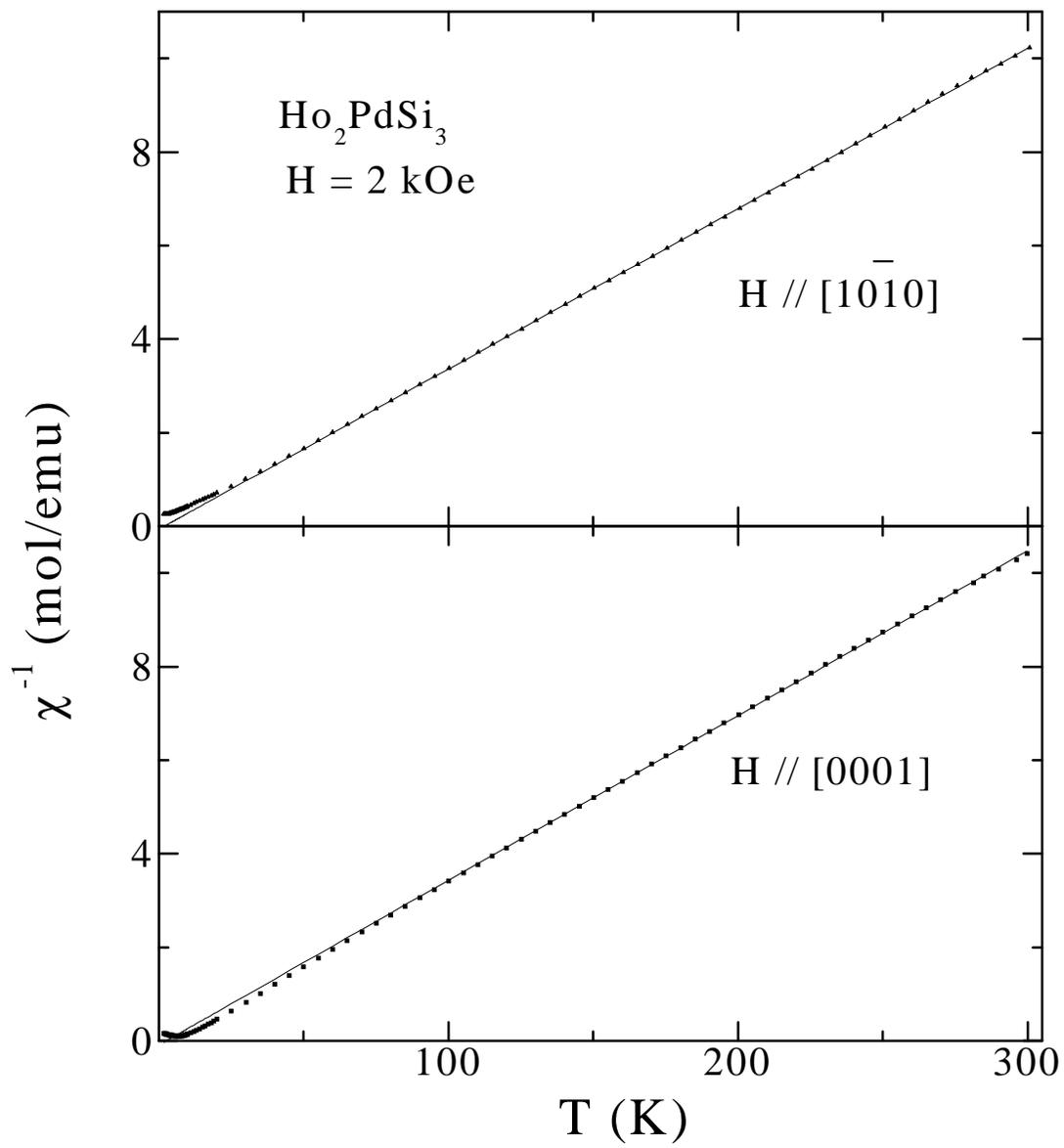
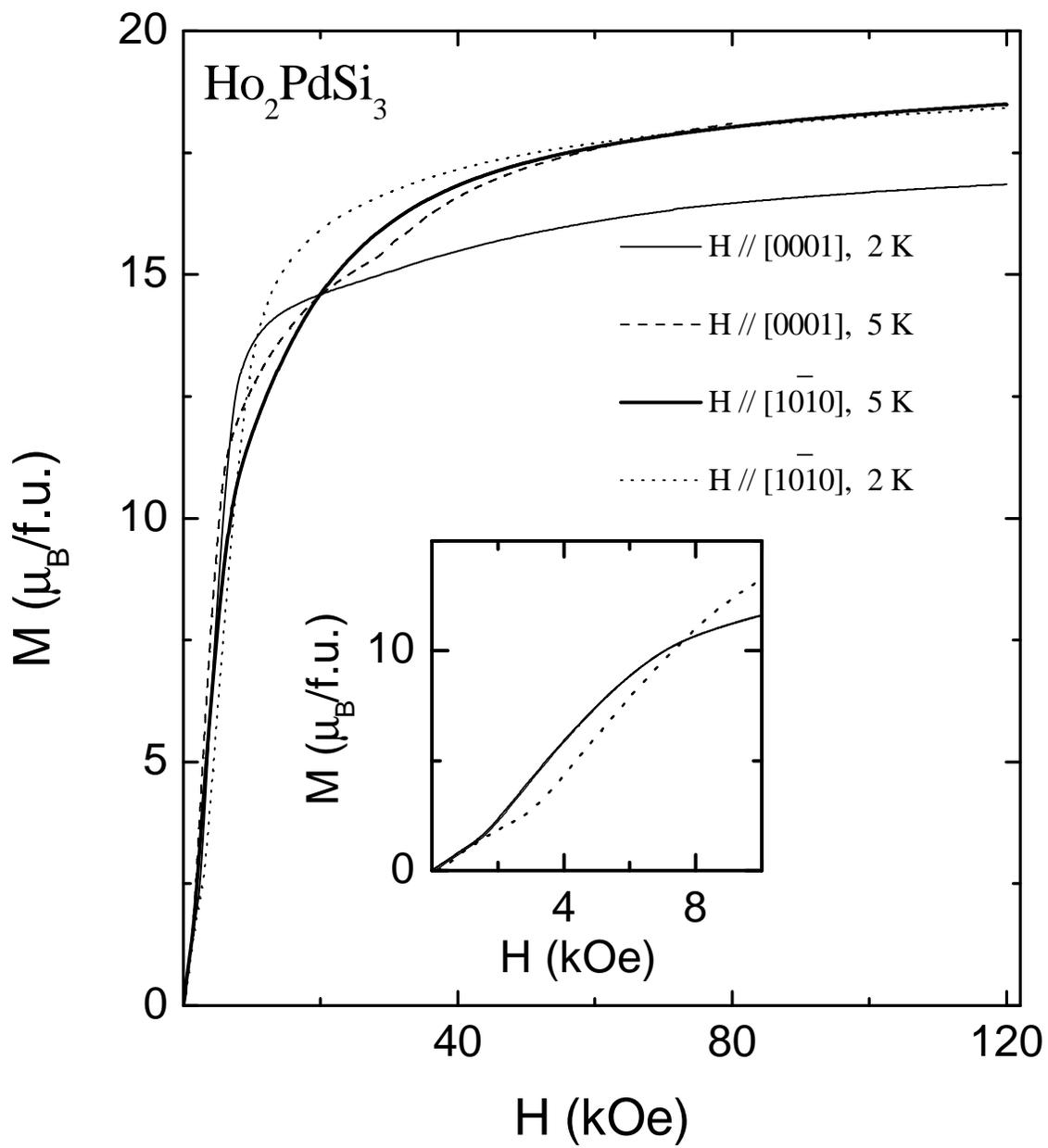
\end{document}